# Finite data-size scaling of clustering in earthquake networks

Sumiyoshi Abe[a,b], Denisse Pastén[c] and Norikazu Suzuki[d]

[a]*Department of Physical Engineering, Mie University, Mie 514-8507, Japan*

[b]*Institut Supérieur des Matériaux et Mécaniques Avancés, 44 F. A. Bartholdi, 72000 Le Mans, France*

[c]*Departamento de Física, Facultad de Ciencias, Universidad de Chile, Casilla 653, Santiago, Chile*

[d]*College of Science and Technology, Nihon University, Chiba 274-8501, Japan*

**ABSTRACT**

Earthquake network is known to be of the small-world type. The values of the network characteristics, however, depend not only on the cell size (i.e., the scale of coarse graining needed for constructing the network) but also on the size of a seismic data set. Here, discovery of a scaling law for the clustering coefficient in terms of the data size, which is refereed to here as *finite data-size scaling*, is reported. Its universality is shown to be supported by the detailed analysis of the data taken from California, Japan and Iran. Effects of setting threshold of magnitude are also discussed.

*PACS:* 89.75.Da, 91.30.Dk, 05.90.+m

*Keywords:* Earthquake networks, Finite data-size scaling, Clustering coefficient

# 1. Introduction

Seismicity has been attracting continuous interest from the viewpoint of complex systems science. Two celebrated classical laws known as the Gutenberg-Richter law [1] and Omori law [2] highlight its complexity: the former tells the scaling relation between frequency of events and released energies (the logarithm of which is magnitude), and the latter states that frequency of aftershocks following a main shock temporally decays as a power law, implying a slow relaxation process. Furthermore, recent studies show that both statistics of spatial distance between the hypocenters of two successive events [3] and statistics of time interval between two successive events [4,5] strongly deviate from Poissonian one. In addition, it is also known [6] that an earthquake can be triggered by the foregoing one that is more than 1000 km away. Thus, the correlation length is divergently large, exhibiting a strong similarity to critical phenomena.

These facts allow us to frame the following working hypothesis: two successive events are indivisibly correlated at least at the statistical level, no matter how large their Euclidean distance is.

In spite of the long tradition of seismological research, microscopic dynamics governing seismicity is still largely unknown, and therefore, it is of central importance to clarify further properties of correlations. In recent works [7], we have introduced the concept of earthquake network in order to reveal complexity of seismicity both qualitatively and quantitatively. We have proposed a method for constructing a growing stochastic network from a seismic data set. There, a single parameter is contained: it is the size of cubic cells to which a geographical region under consideration is divided. Earthquake network constructed is a complex network. It has been shown in Refs. [7-9] that it is of the small-world [10] and scale-free [11] type, hierarchically organized [12], and possesses the assortative mixing property [13]. Its evolution has turned out to characterize a main shock in a peculiar manner [14]. In addition, there exists a scaling relation between the exponent, $\gamma$, of the power-law connectivity distribution [ $P(k) \sim 1/k^{\gamma}$ with $k$ being connectivity] and network spectral density [15].

In more recent works [16,17], we have studied the dependence of the characteristics of earthquake network on the cell size. There, we have found that the values of the exponent, $\gamma$, of the power-law connectivity distribution and the clustering coefficient [10], $C$, (see Sec. 3) come to take the invariant values

$$\gamma \approx 1, \tag{1}$$

$$C \approx 0.85, \tag{2}$$

respectively, if the cell size becomes larger than a certain value, depending on the data set. Universality of these values was confirmed by the analysis of three independent data sets taken from California, Japan and Iran. On the other hand, in the regimes of the small cell size, the value of $C$ in Iran behaves quite differently than those in California and Japan [16,17]. An apparent difference between the former two regions and the latter one is in the data size.

In this paper, we wish to clarify how the data size affects the dependence of the small-world earthquake network on the cell size. Specifically, we focus our attention on the behavior of the clustering coefficient of the network. We shall show, by performing the detailed analysis of the seismic data taken from California, Japan and Iran, that there exists a scaling law in terms of finiteness of the data size, which is referred to here as *finite data-size scaling*. Combined with the result in Eq. (2), this law enables one to determine the cell size for a data set in a geographical region under consideration. We further discuss the effects of setting threshold of magnitude on the scaling.

2. **Construction of earthquake network: Review**

To make the present article self-contained, it seems appropriate to devote this section to a brief review of the method for constructing earthquake network proposed in Ref. [7].

The procedure is as follows. Firstly, we divide a geographical region under consideration into cubic cells. Secondly, we identify a cell with a vertex of a network if earthquakes with any values of magnitude above a certain detection threshold occurred therein. Thirdly, we connect two vertices by an edge, if they are of two successive events. If two successive events occur in the same cell, then we attach a tadpole (i.e., a self-loop) to that vertex. These edges and tadpoles represent the event-event correlations, in conformity with the working hypothesis mentioned in the preceding section, that is, two successive events are statistically correlated no matter how large their Euclidean distance is.

Using this procedure, we can map a given seismic time series to a growing stochastic network, which is an earthquake network that we have been referring to.

We make several comments on the procedure. First of all, it contains a single parameter: the cell size, *L*. Once a set of cells is fixed, then we can unambiguously define an earthquake network for a seismic data set. We note that an earthquake network is a directed network in its nature. Directedness does not matter in statistical analysis of connectivity (or degree, i.e., the number of edges attached to the vertex under consideration), which is needed for examining the scale-free property. This is because the in-degree and out-degree are identical for each vertex except the initial and final vertices in analysis. Thus, vertices except the initial and final ones have the even-number values of connectivity. An important point is that a full directed earthquake network should be reduced to a simple undirected network, when its small-world property is examined (see Fig. 1). That is, we have to remove tadpoles and replace each multiple edge by a single edge.

To practically set up cells and identify a cell for each earthquake, we employ the following natural procedure. Let $\theta_0$ and $\theta_{max}$ be the minimal and maximal values of latitude covered by a data set, respectively. Similarly, let $\phi_0$ and $\phi_{max}$ be the minimal and maximal values of longitude. And let $\theta_{av}$ be the sum of the values of latitude of all the events divided by the number of events contained in the analysis. The hypocenter of the $i$-th event is denoted by $(\theta_i, \phi_i, z_i)$, where $\theta_i$, $\phi_i$ and $z_i$ are the values of latitude, longitude and depth, respectively. The north-south distance between $(\theta_0, \phi_0)$ and $(\theta_i, \phi_i)$ reads $d_i^{NS} = R \cdot (\theta_i - \theta_0)$, where $R (\cong 6370 \text{ km})$ is the radius of the Earth. On the other hand, the east-west distance is given by $d_i^{EW} = R \cdot (\phi_i - \phi_0) \cdot \cos\theta_{av}$. In these expressions, all the angles are described in the unit of radian. The depth is simply $d_i^D = z_i$. Now, starting from the point $(\theta_0, \phi_0, z_0 \equiv 0)$, we divide the region into cubic cells with a given value of the cell size, $L$. Using $d_i^{NS}$, $d_i^{EW}$ and $d_i^D$, we can identify the cell of the $i$-th event.

### 3. Finite data-size scaling of clustering coefficient

Now, let us address ourselves to studying the clustering property of earthquake network. A network to be considered is a simple network reduced from a full network, as mentioned in Sec. 2.

The clustering coefficient, $C$, of a simple network with $N$ vertices is given by [10]

$$C = \frac{1}{N} \sum_{i=1}^{N} c_i, \tag{3}$$

where

$$c_i = \frac{\text{(number of edges connecting neighboring vertices of } i\text{-th vertex)}}{k_i(k_i-1)/2} \qquad (4)$$

with $k_i$ being connectivity of the $i$-th vertex. Let $A$ be the adjacency matrix of a simple network. Its element $(A)_{ij}$ is 1 (0), if the $i$-th and $j$-th vertices are connected (unconnected) by an edge, and $(A)_{ii}=0$. Using this matrix, $c_i$ is written as follows:

$$c_i = \frac{e_i}{e_{i,\max}}, \qquad (5)$$

where

$$e_i = (A^3)_{ii}, \qquad (6)$$

and $e_{i,\max} = k_i(k_i-1)/2$ is the maximum value of $e_i$, which is realized when all the neighboring vertices of the $i$-th vertex are connected each other.

The value of the clustering coefficient of a small-world network is much larger than that of a completely random network [10]. It is known [7-9] that this is indeed the case for earthquake network.

Like other characteristics of a network, the clustering coefficient is dimensionless. In the case of earthquake network, however, its numerical value depends on the cell size, $L$, of division of a geographical region. Accordingly, the dimensionless cell size should be considered instead of $L$ itself. There are two candidates for it:

$$l_3 \equiv L / (L_{LAT} L_{LON} L_{DEP})^{1/3}, \qquad (7)$$

$$l_2 \equiv L / (L_{LAT} L_{LON})^{1/2}, \qquad (8)$$

where $L_{LAT}$, $L_{LON}$ and $L_{DEP}$ are the dimensions of the whole region under consideration in the directions of latitude, longitude and depth, respectively. The implications of $l_2$ and $l_3$ are discussed in Refs. [16,17] in view of the nature of seismicity. At first glance, $l_3$ seems natural. However, there is an empirical fact that in California, Japan and Iran, the earthquake networks are quite two-dimensional, since the majorities of events occur in the shallow regions there (see the later discussion). Therefore, we examine both of them in the present work.

The value of the clustering coefficient of earthquake network depends on both the dimensionless cell size and the data size, $n$, which is the number of events contained in a data set to be analyzed. That is,

$$C = C(l_\alpha, n) \qquad (\alpha = 2, 3). \qquad (9)$$

In what follows, we shall see that this quantity possesses a remarkable property.

The data sets we employ here are those taken from (i) California; http://www.data.scec.org, (ii) Japan; http://www.hinet.bosai.go.jp, and (iii) Iran; http://irsc.ut.ac.ir/. The periods and the geographical regions covered are as follows: (i) between 00:25:8.58 on January 1, 1984 and 23:15:43.75 on December 31, 2006, 28.00°N–39.41°N latitude, 112.10°W–123.62°W longitude with the maximal depth

175.99 km, (ii) between 00:02:29.62 on June 3, 2002 and 23:54:36.21 on August 15, 2007, 17.96°N–49.31°N latitude, 120.12°E–156.05°E longitude with the maximal depth 681.00 km, and (iii) between 03:08:11.10 on January 1, 2006 and 18:26:21.90 on December 31, 2008, 23.89°N–43.51°N latitude, 41.32°E–68.93°E longitude with the maximal depth 36.00 km. Accordingly, the values of $(L_{\text{LAT}}L_{\text{LON}}L_{\text{DEP}})^{1/3}$ and $(L_{\text{LAT}}L_{\text{LON}})^{1/2}$ in Eqs. (7) and (8) are respectively as follows: (i) 617.80 km and 1157.51 km (ii) 1973.78 km and 3360.26 km and (iii) 583.45 km and 2348.86 km. Since these are simply rescaling factors to make cell size dimensionless, we shall commonly employ their values throughout the present work even when we remove events from the data to change data size. The total numbers of events contained in these periods are (i) 404106, (ii) 681546, and (iii) 22845. The majorities of events are shallow: 90 % of the events are shallower than (i) 13.86 km, (ii) 69.00 km, (iii) 26.60 km. This fact makes it meaningful to consider also $l_2$ in Eq. (8).

We have examined the physical property of the clustering coefficient in Eq. (9) for these three independent data sets. In Fig. 2, where no threshold is set on the values of magnitude (for thresholding, see Section 5), we present the plots of $C$ with respect to $l_3$ for several different values of the data size, $n$. (We could consider only three different values of $n$ for the data in Iran, since the total number of events contained in it is quite small, compared to those in California and Japan.) One sees a general tendency that, for each value of $l_3$, the larger the data size is, the larger the value of $C$ is. This is natural, since a larger data tends to make a network closer to a complete one that has the maximum value, $C=1$. Although the clustering coefficient approaches its universal value, 0.85, for large cell size in all three regions, the Iranian one behaves differently from the other two for small cell size: that is, the universal value is approached slower in Iran. Fig. 2 shows that this difference is due to smallness of the data size in Iran (see also Ref. [16]).

Now, our main discovery is that, as can be seen in Fig. 3, the data collapses are nicely realized if the following scaling is made:

$$C = \tilde{C}(l_\alpha / f(n)) \qquad (\alpha = 2, 3), \tag{10}$$

where $\tilde{C}$ is a scaling function and $f(n)$ chosen here is of the form

$$f(n) = 1 + \left(\frac{n_0}{n}\right)^a \tag{11}$$

with $n_0$ and $a$ being constants. Quite remarkably, in Fig. 3, the *same fixed values*, $n_0 = 1 \times 10^5$ and $a = 0.43$ for both $l_3$ and $l_2$, are employed for these data collapses in all three geographical regions. The law in Eq. (10) describes the finite data-size scaling that we have been referring to.

4. **Cell size as scale of coarse graining**

The scaling law presented in the preceding section allows one to determine the scale of coarse graining (i.e., the cell size) depending on the data size. Fig. 3 shows that the clustering coefficient in Eq. (10) takes the invariant value

$$\tilde{C}(l_\alpha / f(n)) \approx 0.85 \qquad (\alpha = 2, 3), \tag{12}$$

if the cell size becomes larger than a certain value, $l_\alpha^*$, which fixes the scale of coarse graining. In both California and Japan, $l_3^* / f(n)$ is about 0.04, whereas it is roughly 0.10 in Iran. On the other hand, $l_2^* / f(n)$ is about 0.015 for California and Japan, whereas about 0.025 in Iran. Such a numerical discrepancy comes from the size of the whole data set. An important point is that the scaling functions in California and Japan well coincide with each other. On the other hand, the scaling function in Iran has a shape different from them. This implies that the size of the Iranian data is still too small, whereas the size of the Californian and Japanese data may already be close to the "thermodynamic limit".

Finally, we make a comment on a possible physical interpretation of the saturation of the scaling function with respect to the cell size. As the cell size increases, the number of vertices decreases, and the network tends to approach a complete graph, i.e., a fully linked network, with the maximum value of the clustering coefficient ($C = 1$). This effect increases the value of $C$. On the other hand, however, larger cells "swallow" more triangular loops attached to them, as can be seen from the $A^3$ – nature of $C$ in Eq. (6). This effect decreases the value of $C$. Disappearance of the cell-size dependence of the clustering coefficient can be understood as compensation of these two competing effects.

**5.    Comment on effects of threshold of magnitude**

There is yet another possibility of changing data size. It is to set threshold of magnitude. In this section, we discuss this issue and show that the finite data-size scaling is valid also in this case.

Here, we have analyzed only the data sets of California and Japan, since the size of the full Iranian data is too small.

In Fig. 4, we show how the cell-size dependence of the clustering coefficient is altered by thresholding. The trend is quite similar to that in Fig. 2. In fact, as can be seen in Fig. 5, the scaling in Eq. (10) turns out to hold well in both California and Japan. The value of $n_0$ in Eq. (11) is 1000 times larger here, i.e., $n_0 = 1 \times 10^8$. However, quite remarkably, the exponent $a$ remains unchanged as $a = 0.43$, indicating its universality.

**6.    Concluding remarks**

We have studied the clustering structure of earthquake network and examined its dependence on the size of a real seismic data. We have discovered that there exists a scaling law for the clustering coefficient in terms of the cell size needed for constructing a network and the data size. We have ascertained universality of this concept by employing three independent data taken from California, Japan, and Iran. We have found that the scaling function associated with the clustering coefficient approaches a universal invariant value in Eq. (12), if the cell size becomes larger than a certain value. These results, in turn, allow one to fix the cell size with reference to the data size. We have also discussed the effects of setting threshold of magnitude on the scaling and have found that the exponent [$a$ in Eq. (11)] remains unchanged.

We wish to make the following additional comment on unavoidable incompleteness of a seismic data set. Naively, one might wonder if the incompleteness changes the results obtained here. Regarding this point, it should be noted that a complex network is strongly tolerant against "random attacks", i.e., random removals of vertices [18]. Earthquake network does not have the centralities with small values of connectivity. Since incompleteness of a data set is not biased (i.e., not due to "intelligent attacks"), we can confidently believe robustness of the present results.

*Note added*. In this work, we have employed the fixed values of the rescaling factors $(L_{LAT} L_{LON} L_{DEP})^{1/3}$ and $(L_{LAT} L_{LON})^{1/2}$ for $l_3$ and $l_2$ in Eqs. (7) and (8) for each geographical region. When the data size is changed, these values also change, in general. Actually, we have examined this point and confirmed that the scaling reported in this work is valid even if the above-mentioned values alter according to the change of the data size.

**Acknowledgements**

S. A. was supported in part by a Grant-in-Aid for Scientific Research from the Japan Society for the Promotion of Science. D. P. acknowledges the scholarship of CONICyT and the hospitality of Mie University extended to her.**References**

[1]  B. Gutenberg, C.F. Richter, Seismicity of the Earth and Associated Phenomena, Princeton University Press, Princeton, 1954.

[2]  F. Omori, J. Coll. Sci. Imp. Univ. Tokyo 7 (1894) 111.

[3]  S. Abe, N. Suzuki, J. Geophys. Res. 108(B2) (2003) 2113, ESE 19.

[4]  Á. Corral, Phys. Rev. Lett. 92 (2004) 108501.

[5]  S. Abe, N. Suzuki, Physica A 350 (2005) 588.

[6] D.W. Steeples, D.D. Steeples, Bull. Seismol. Soc. Am. 86 (1996) 921.

[7]  S. Abe and N. Suzuki, Europhys. Lett. 65 (2004) 581. See also, S. Abe, N. Suzuki, Nonlin. Processes Geophys. 13 (2006) 145; in Encyclopedia of Complexity and Systems Science, edited by R.A. Meyers, pp. 2530, Springer, New York, 2009.

[8]  S. Abe, N. Suzuki, Physica A 337 (2004) 357.

[9]  S. Abe, N. Suzuki, Phys. Rev. E 74 (2006) 026113.

[10]  D.J. Watts, S.H. Strogatz, Nature (London) 393 (1998) 440.14

# Figure Captions

**Fig. 1**  Schematic descriptions of earthquake networks. (A) Full directed network, and (B) undirected simple network reduced from (A). The dotted lines describe the initial and final events contained in analysis.

**Fig. 2**  Plots of the clustering coefficient with respect to (a) $l_3$ and (b) $l_2$ for several values of the data size, $n$. (i) California: $\triangle, +, \square, \times, \bigcirc$, and $*$ are the first 5000, 10000, 50000, 100000, 200000, and 404106 events in the data set, respectively. (ii) Japan: $\triangle, +, \square, \times, \bigcirc$, and $*$ are the first 5000, 10000, 50000, 100000, 200000, and 681546 events in the data set, respectively. (iii) Iran: $\times, \bigcirc$, and $*$ are the first 5000, 10000, and 22845 events in the data set, respectively. All quantities are dimensionless.

**Fig. 3** Collapses of the data in Fig. 2. In all of these three cases, the same fixed values, $n_0 = 1 \times 10^5$ and $a = 0.43$, are employed in Eq. (11). All quantities are dimensionless.

**Fig. 4** Plots of the clustering coefficient with respect to (a) $l_3$ and (b) $l_2$ for several values of threshold of magnitude, $M_{th}$: that is, the events having magnitude $M \geq M_{th}$ are included in the analyses. No threshold ($*$), $M_{th}$ = 1.0 (○), 1.5 (×), 2.0 (□), 2.5 (+) and 3.0 (△). The corresponding number s of events included are (i) in California, 404106, 334207, 219836, 105653, 39214 and 11688, and (ii) in Japan, 681546, 342507, 201117, 114941, 62554 and 30907, respectively. All quantities are dimensionless.

**Fig. 5** Collapses of the data in Fig. 4. In both California and Japan, the same values, $n_0 = 1 \times 10^8$ and $a = 0.43$, are employed in Eq. (11) for the scaling in the same form as in Eq. (10). Note that the value of *a* is the same as that in Fig. 3. All quantities are dimensionless.

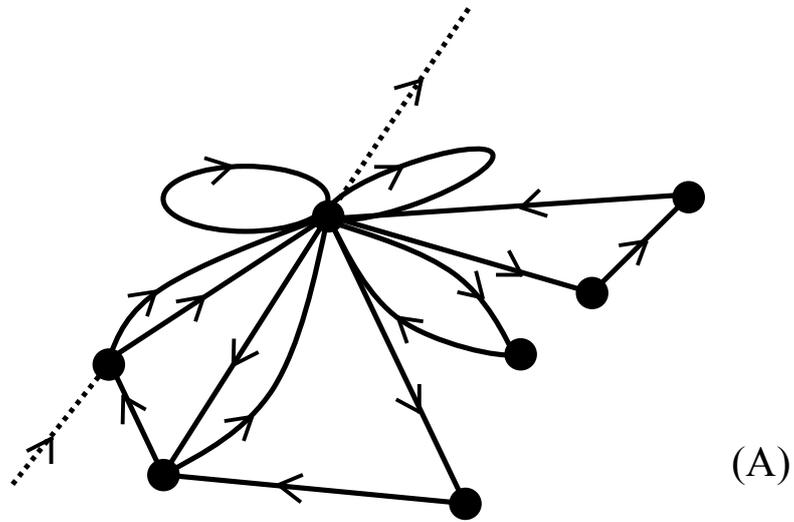

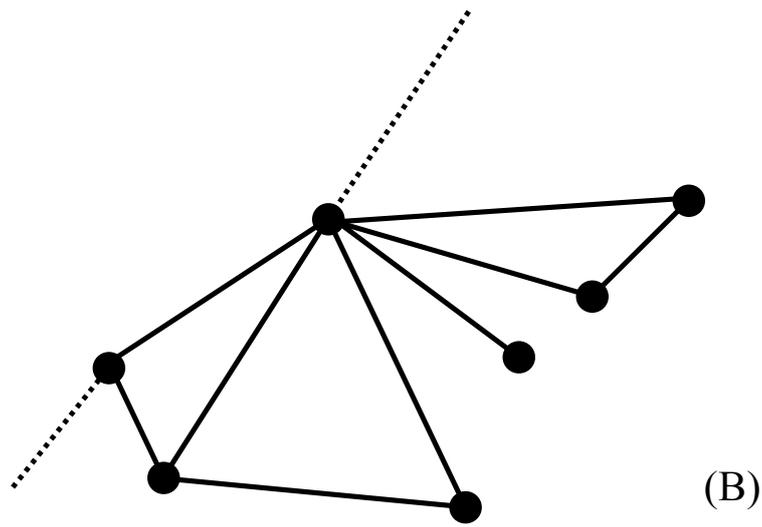

Fig. 1

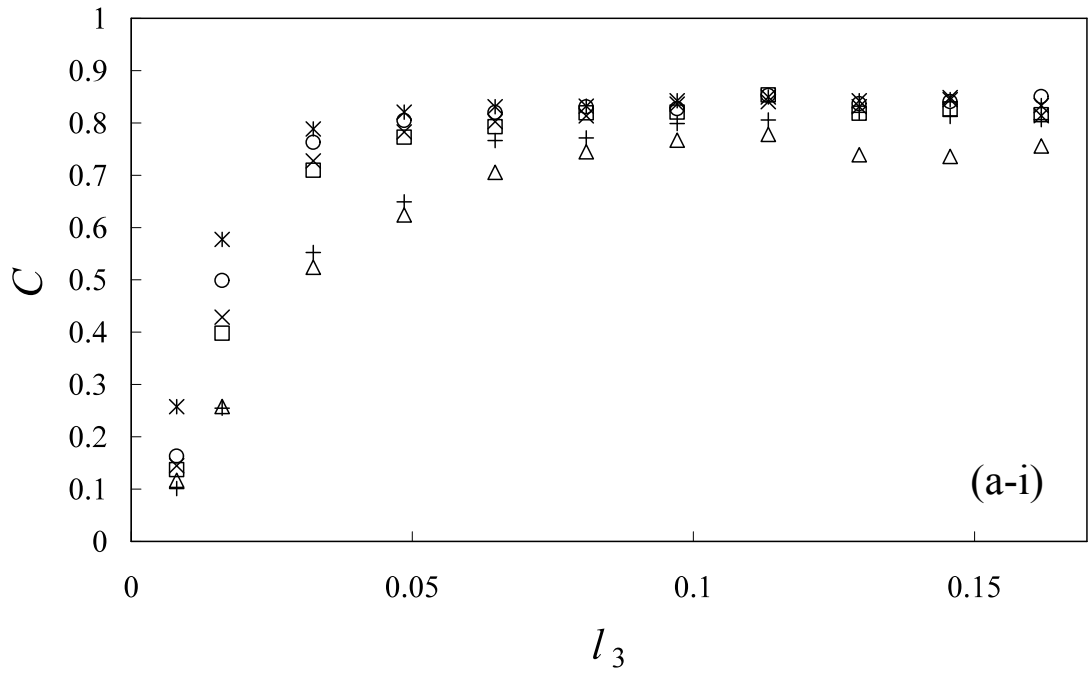

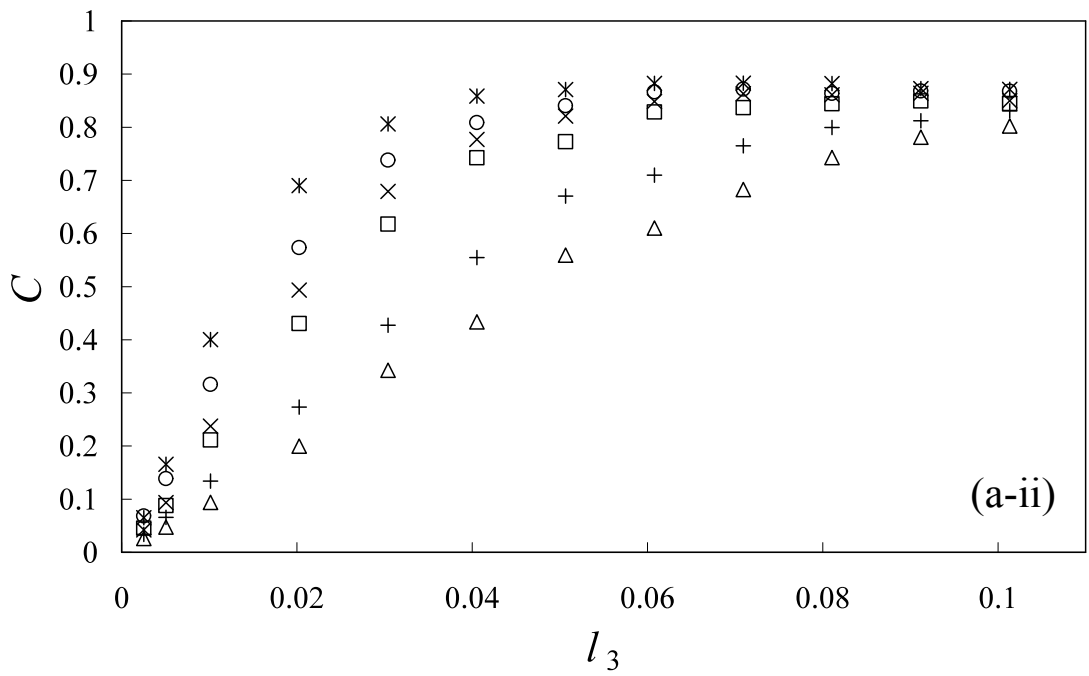

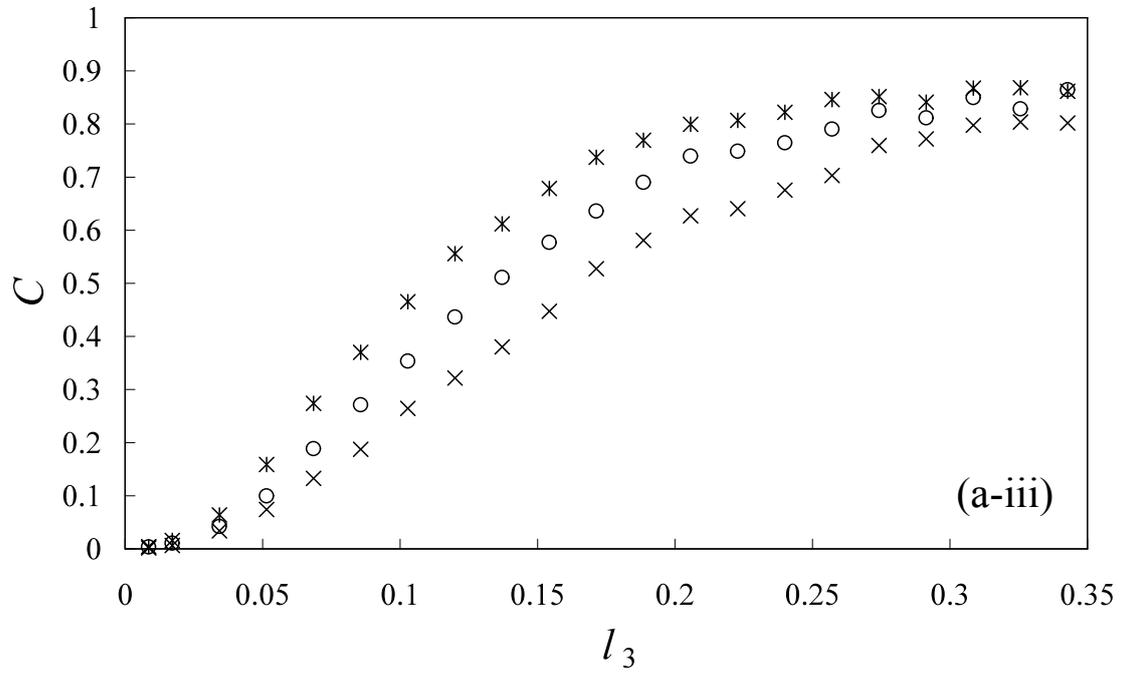

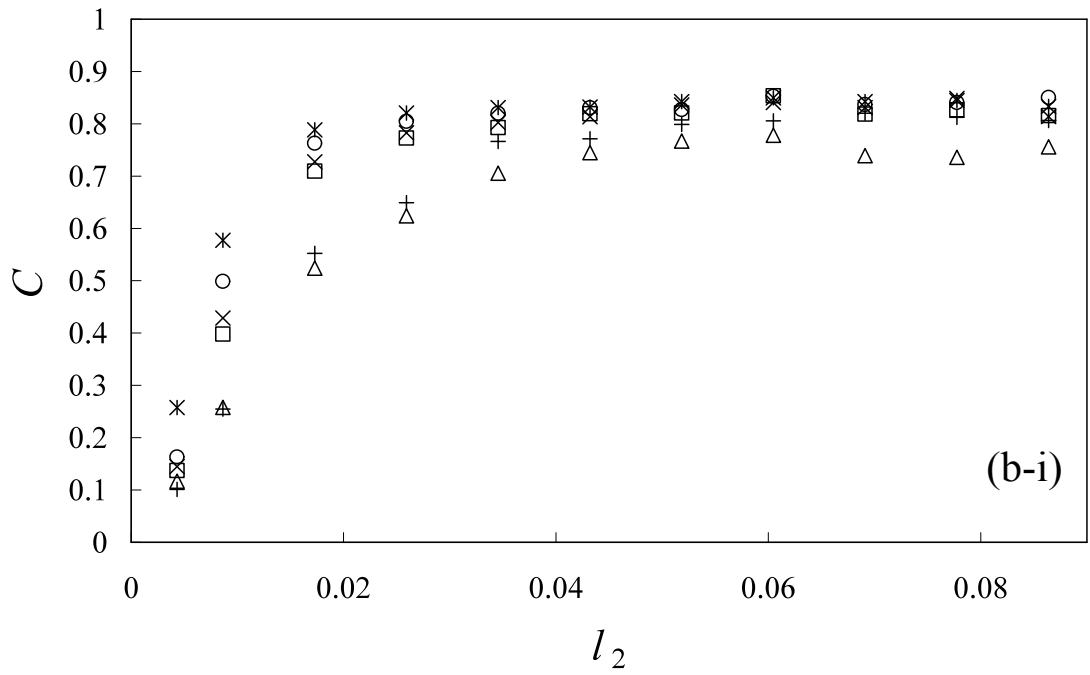

(b-i)

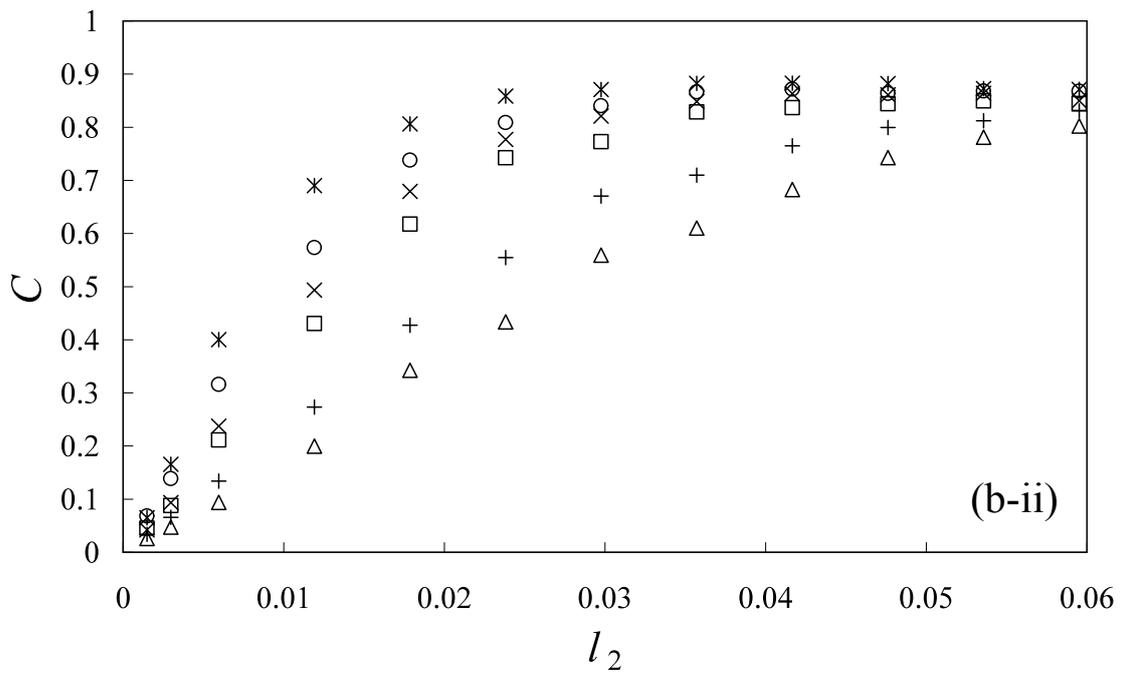

(b-ii)

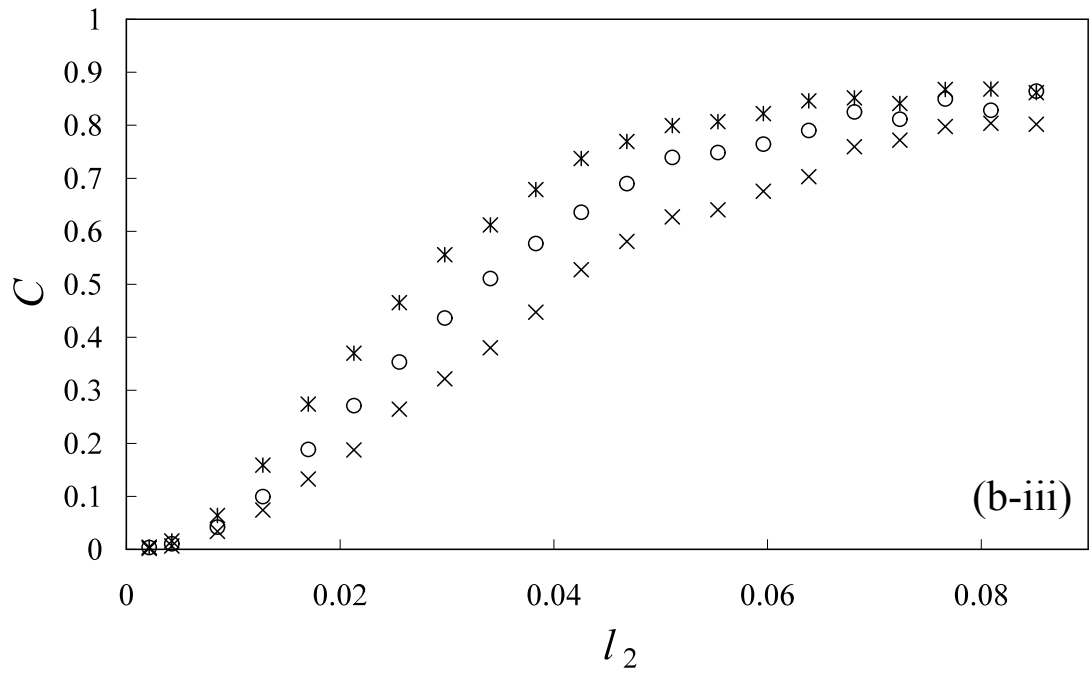

Fig. 2

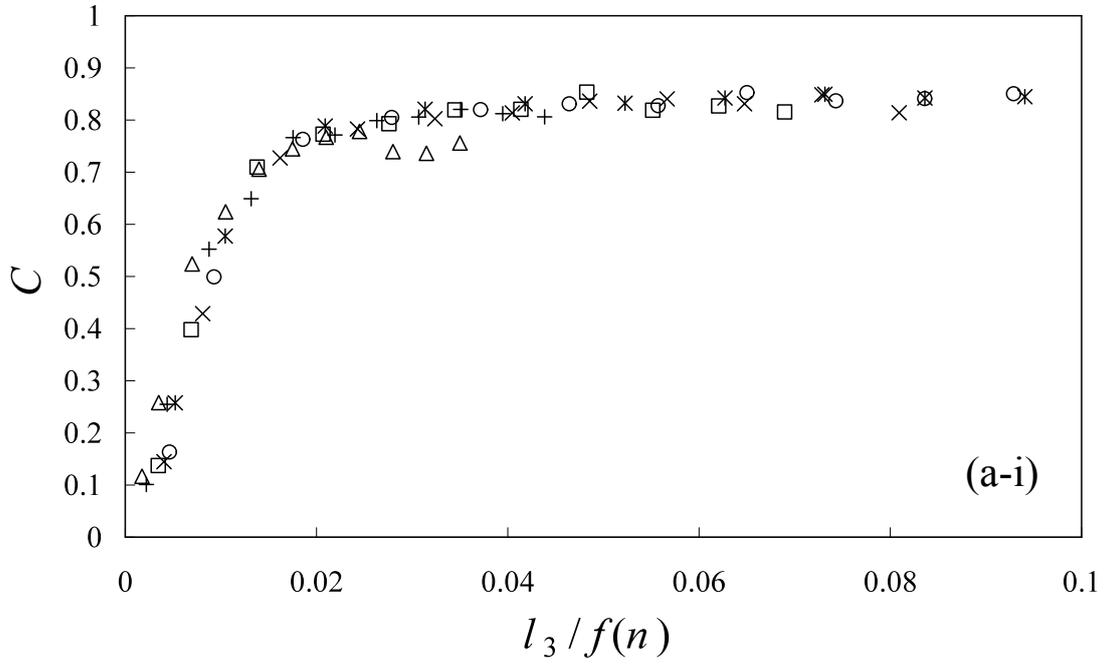

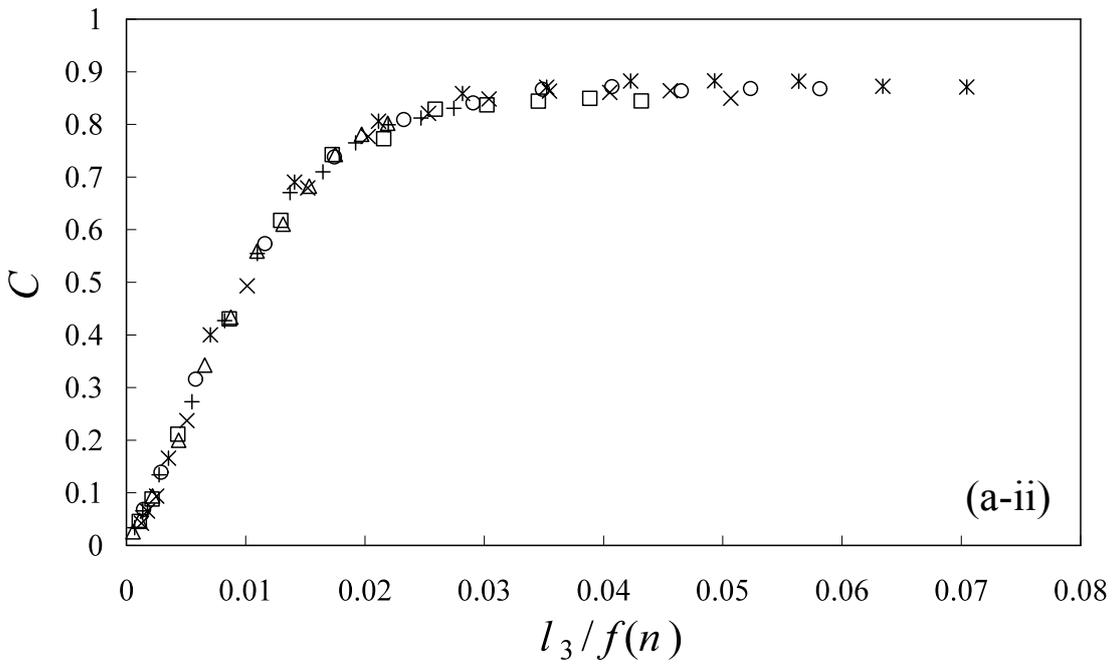

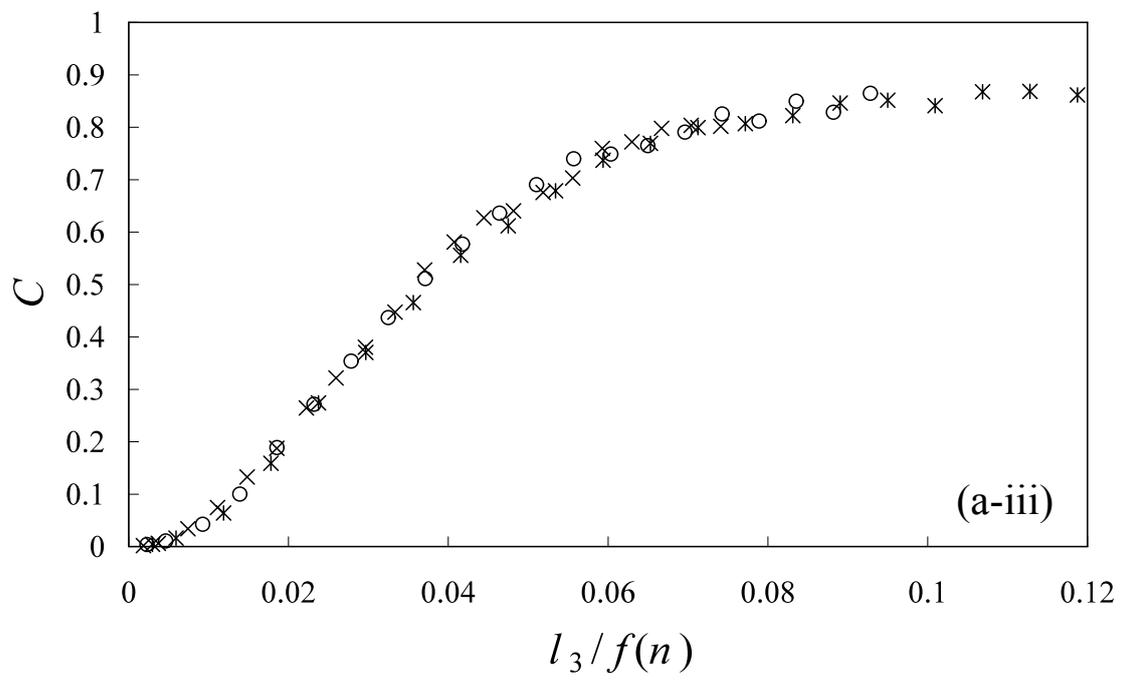

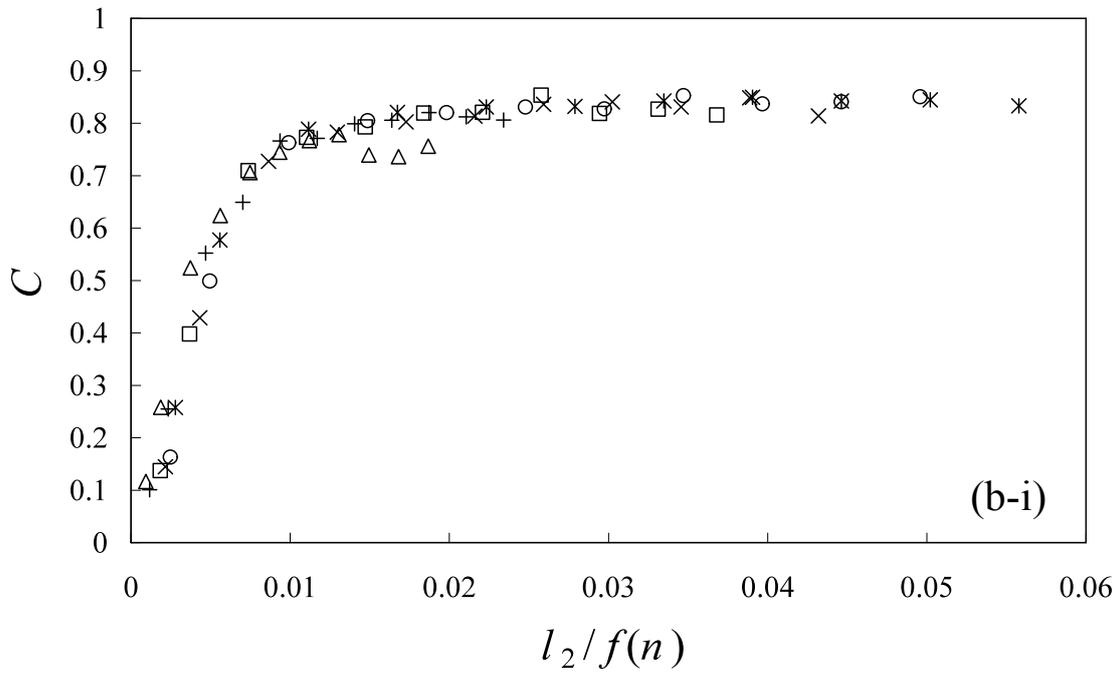

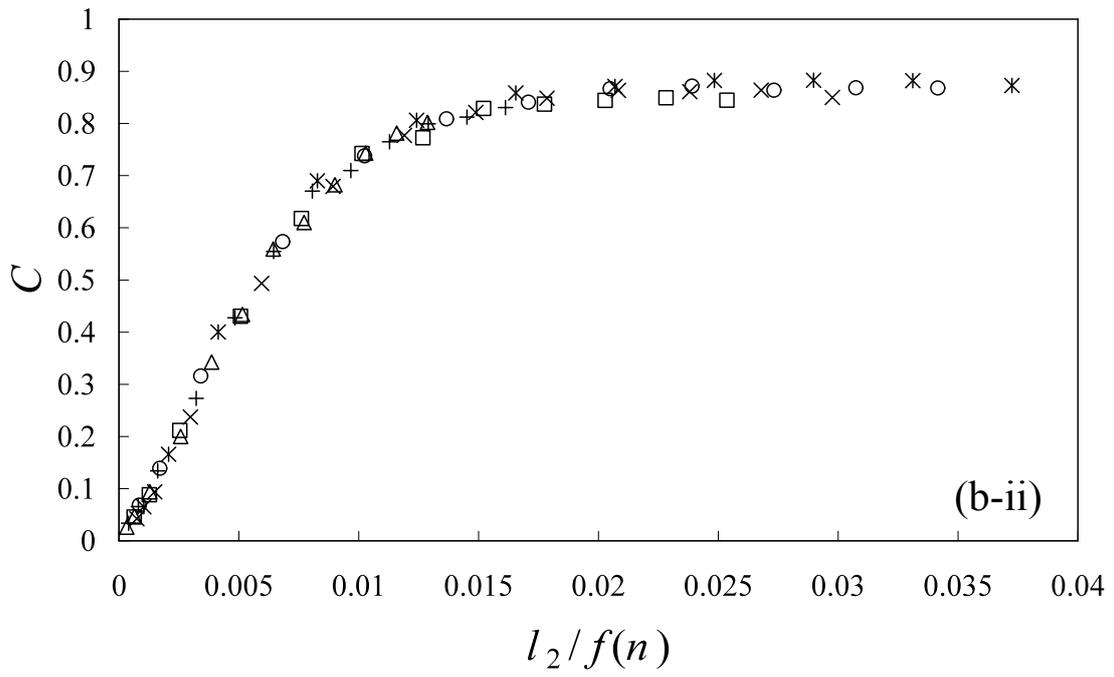

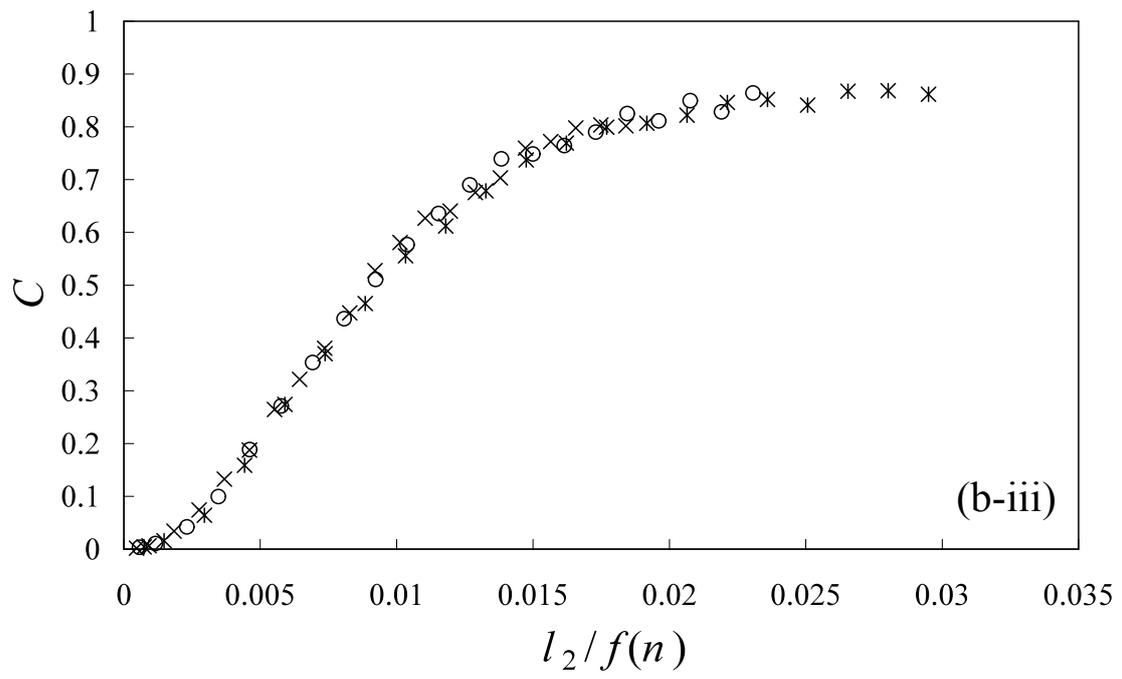

Fig. 3

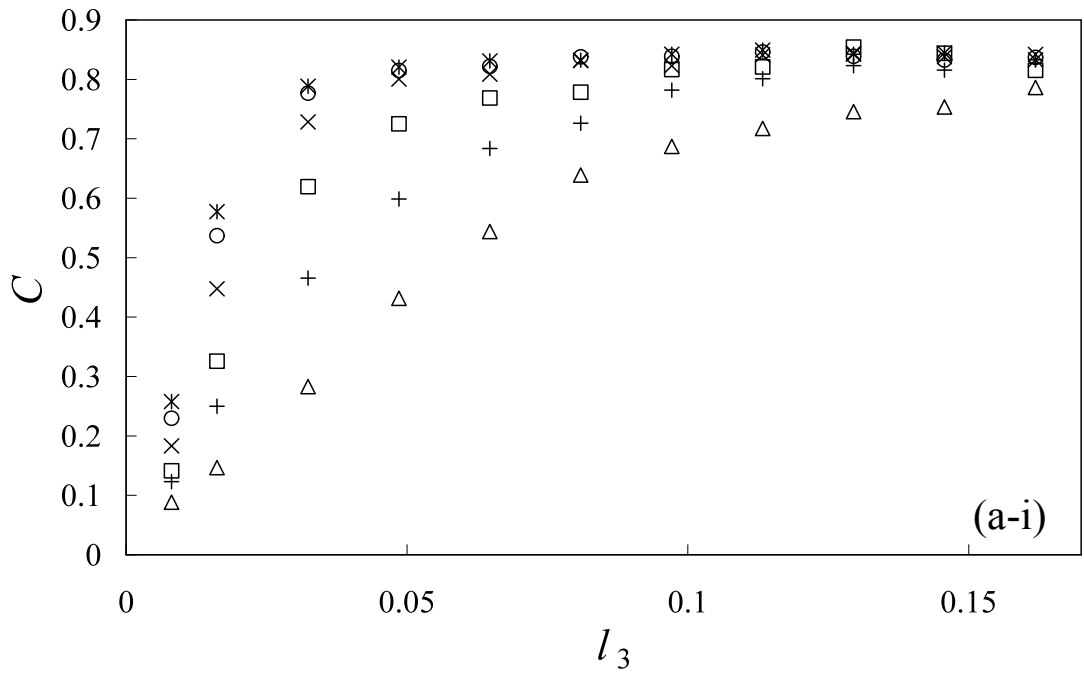

(a-i)

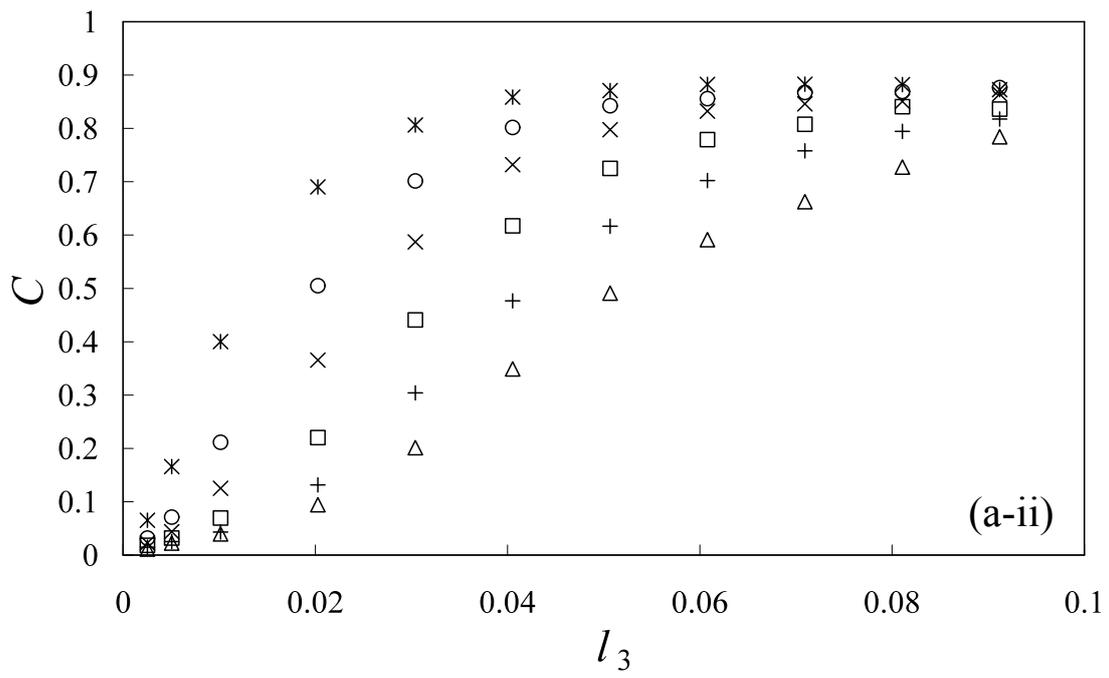

(a-ii)

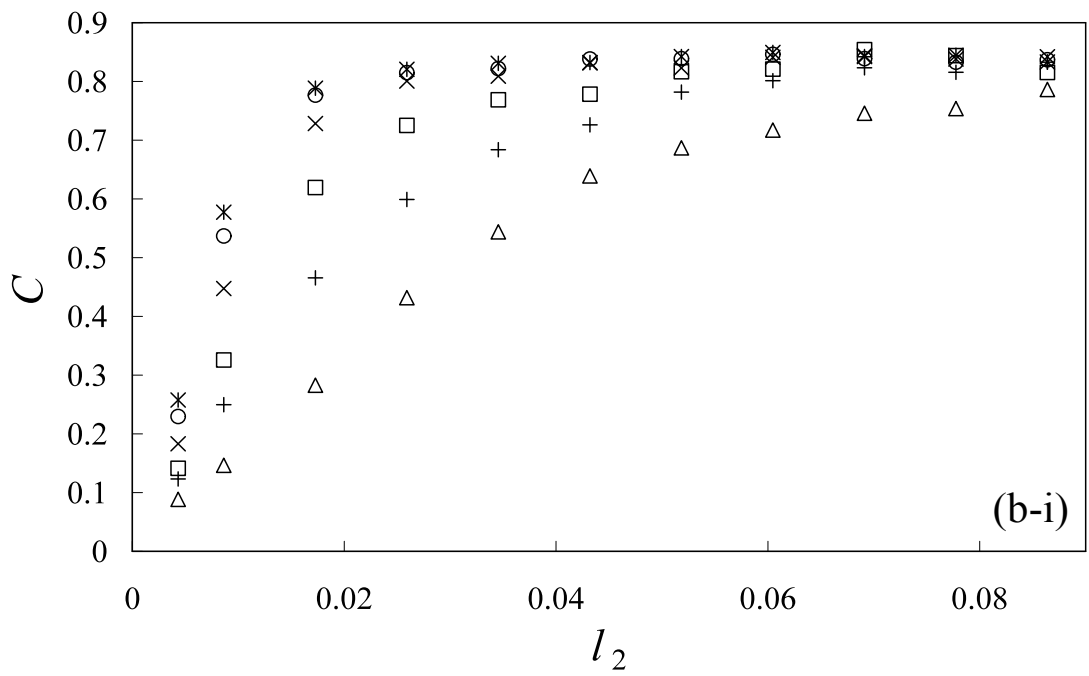

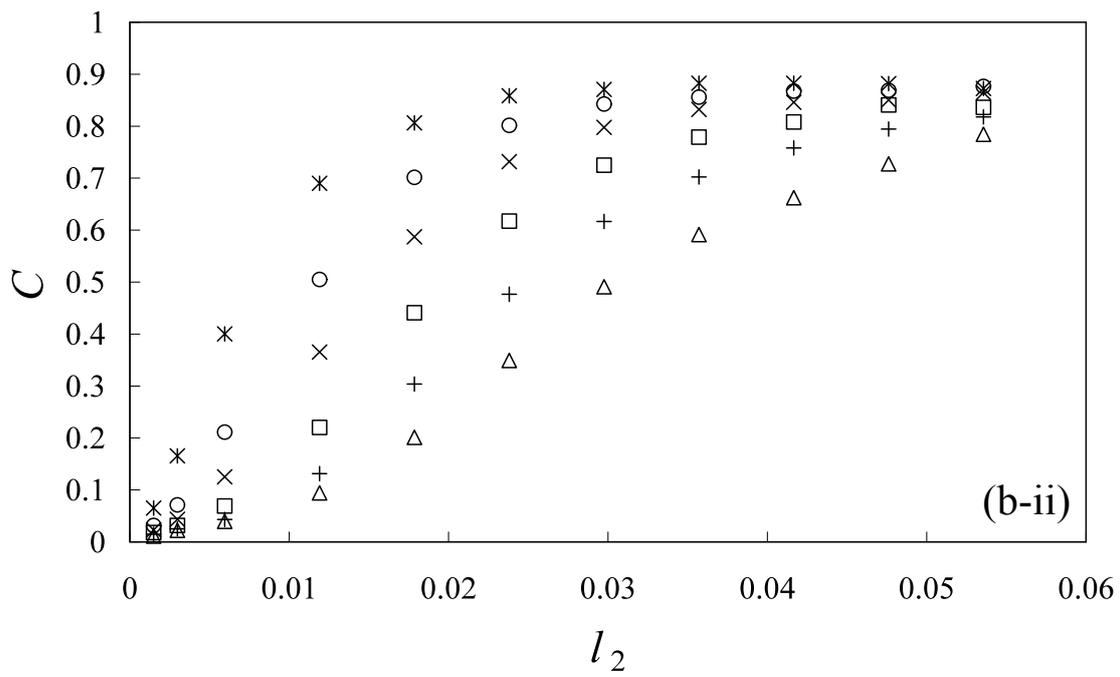

Fig. 4

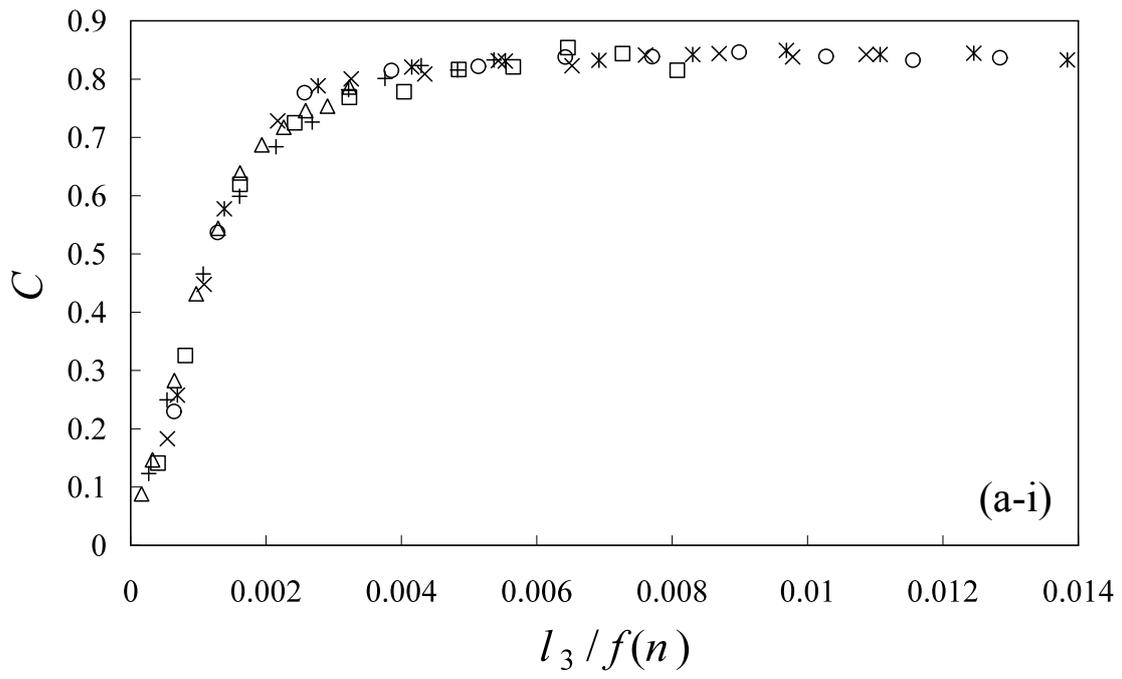

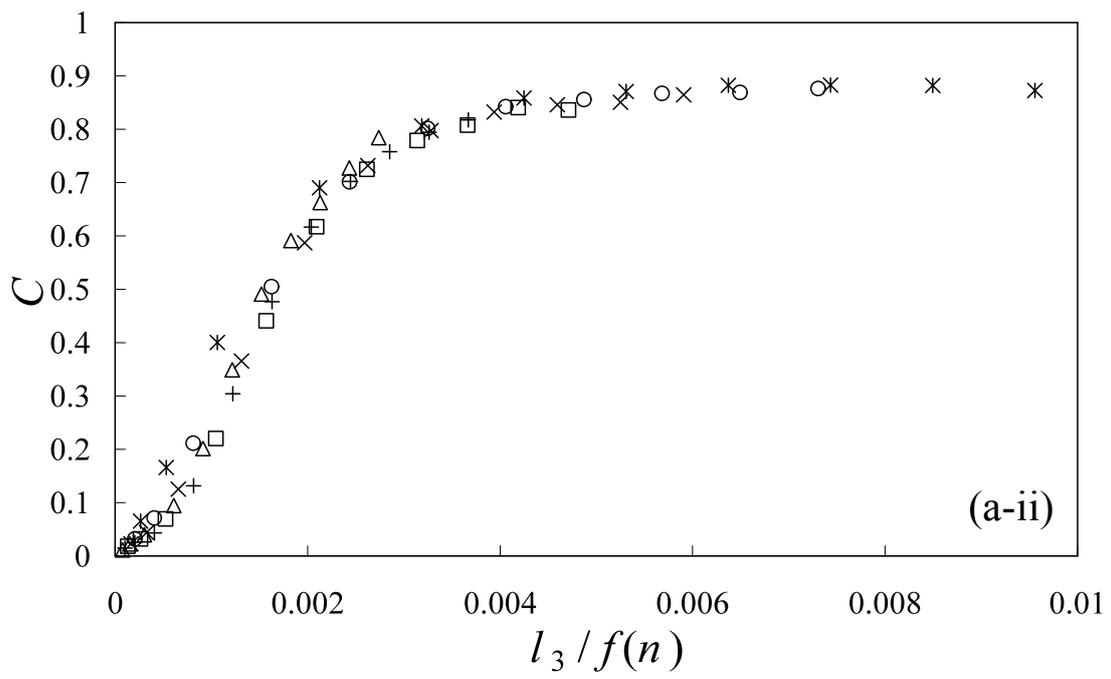

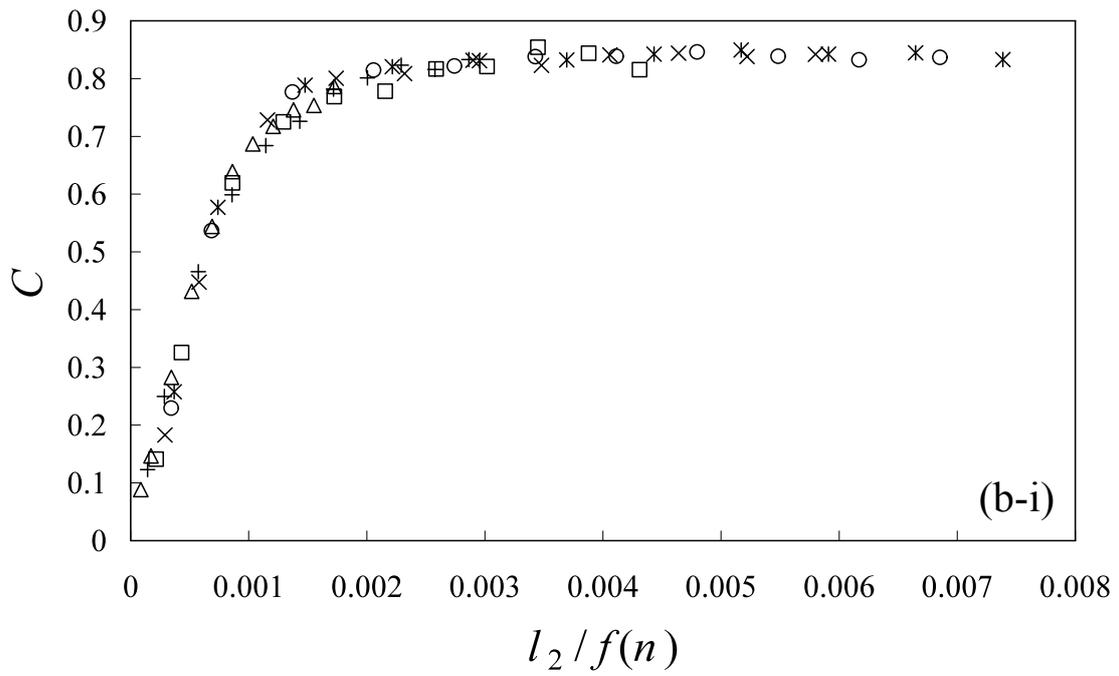

(b-i)

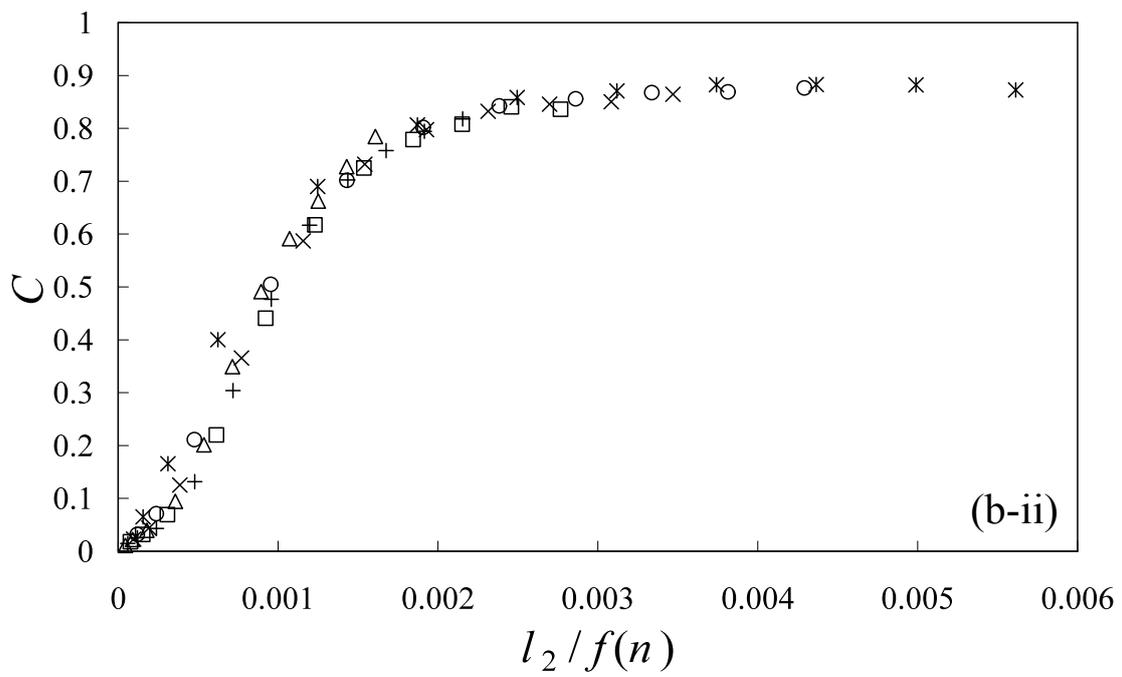

(b-ii)

Fig. 5